\documentclass[11pt]{article}

\usepackage{anysize}
\marginsize{1in}{1in}{1in}{1in}

\usepackage{amssymb}
\usepackage{times,mathptm}
\usepackage{color}

\usepackage[colorlinks]{hyperref} 
\usepackage{graphicx}

\newcommand{\ket}[1]{|#1\rangle}
\newcommand{\bra}[1]{\langle#1|}

\newcommand{\dpo}[1]{#1}

\definecolor{plum}{rgb}{0.45,0,.66}

\title{Bounds on Effective Hamiltonians for Stabilizer Codes}
\author{
\begin{tabular}{lcr} 
Stephen S. Bullock & \quad \quad   & Dianne P. O'Leary\footnote{This
author was supported in part
by the National Science Foundation under Grant CCF 0514213.} \\
Center for Computing Sciences & & Department of Computer Science \\
Institute for Defense Analyses & & University of Maryland \\
& & MCSD, Division 891 \\
& & N.I.S.T. \\
ssbullo@super.org & & oleary@cs.umd.edu \\
\end{tabular}
}
\date{February 4$^{\mbox{\footnotesize th}}$, 2008}

\begin{document}

\maketitle

\begin{abstract}
This \dpo{manuscript} introduces various notions of $k$-locality
of stabilizer codes inherited from the associated stabilizer
groups.  A choice of generators for the group leads to a Hamiltonian
with the code in its groundspace, while a Hamiltonian holding
the code in its groundspace might be called \emph{effective}
if its locality is less than that of a natural choice of generators
(or any choice).  This paper establishes some conditions under which
effective Hamiltonians for stabilizer codes do not exist.  
Our results simplify in the cases of Calderbank-Shor-Steane
stabilizer codes and topologically-ordered stabilizer
codes arising from surface cellulations.
\end{abstract}


\section{Introduction}

\dpo{A simple realization of a Hamiltonian can be achieved
if the Hamiltonian is $k$-local for a small integer $k$.
For adiabatic quantum computing, a Hamiltonian is of interest
because its ground state reveals the answer to an interesting
problem.
We study in this paper the problem of determining lower bounds on the
locality of Hamiltonians whose groundstate is a stabilizer code,
and show that all such Hamiltonians must be at least as complicated as the
underlying stabilizer group.
}

Consider a collection of $n$ qubits
evolving under a constant Hamiltonian $H$.  Write
$\mathcal{H}_1 = \mathbb{C} \{ \ket{0}\} \oplus \mathbb{C}\{\ket{1}\}$
and the $n$-qubit Hilbert space as
$\mathcal{H}_n = (\mathcal{H}_1)^{\otimes n} \cong
\oplus_{j=0}^{2^n-1}\mathbb{C} \{\ket{j}\}$,
so that we might view
$H=\sum_{j,k=0}^{2^n-1} h_{jk}\ket{j}\bra{k} \in
\mathbb{C}^{2^n \times 2^n}$ as a Hermitian matrix.  The notion
of $k$-locality \cite{gadget} has been introduced to estimate
how physically plausible such a Hamiltonian $H$ might be.
To describe this, let $J$ be an $n$-long list of elements of
$\{0,x,y,z\}$.  For such an $J=j_1 j_2 \ldots j_n$, we create
an abbreviation $\sigma_{\otimes J} = \sigma_{j_1} \otimes 
\sigma_{j_2} \otimes \cdots \otimes \sigma_{j_n}$ for the
appropriate tensor product of Pauli matrices.  Let $\mathcal{J}$
denote the set of all such indices $J$. We 
use $\mathcal{H}(2^n)$ to denote the vector space of Hermitian
matrices.  This sets notation for the \dpo{equation}
\begin{equation}
\label{eq:Rbbcombo}
\mathcal{H}(2^n) \ = \ \bigoplus_{J \in \mathcal{J}}
\mathbb{R} \; \{ \sigma_{\otimes J} \} .
\end{equation}
Containment of the right-hand side follows since tensors
of Hermitian matrices are Hermitian, while the equality
follows from linear independence given that the Pauli-tensors
are orthogonal in the matrix inner product 
$A \bullet B = \mbox{Trace}( A \overline{B}^T)$
for $A, B \in \mathbb{C}^{2^n \times 2^n}$. 
Thus, $H$ may also be written \dpo{as} 
\begin{equation}
\label{eq:sumofpaulis}
H \ = \ \sum_{J \in \mathcal{J}} t_J \sigma_{\otimes J},
\quad t_J \in \mathbb{R} .
\end{equation}
If we use $\#J$ to denote the number of nonzero indices, then any
summand $t_J \sigma_{\otimes J}$ of $H$ denotes a $\#J$-body
interaction among the qubits.  We say $H$ is $k$-local when
$k \geq \mbox{max} \{ \# J \; | \; t_J \neq 0 \}$.

Much recent work in quantum complexity theory
considers the ground states of $k$-local Hamiltonians.  
For example, an adiabatic quantum computer
\cite{farhi,gadget} must remain
in the ground-state of a $k$-local Hamiltonian at all times.  
Early works on anyonic excitations of topologically ordered
Hamiltonians \cite{Kitaev,FreedmanMeyer} used Hamiltonians whose
addends were based on the local structure of some lattice.  These
were usually $k$-local for $k$ small.
Square lattices produce four local Hamiltonians
while triangular lattices and their dual hexagonal lattices each
produce six local Hamiltonians.
Another recent topic considers realizing graph states 
as groundstates \cite{graph}.  Realizing a graph
state in this way is of interest since (i) the graph state 
is a \emph{nondegenerate} groundstate which in principle could
be obtained from the physical system by cooling, and (ii)
any quantum circuit may be emulated using one-qubit
rotations and measurements of the graph state
\cite{biggraph}.  Thus realizing a Hamiltonian
for a large enough graph state, cooling the system,
and then applying local control and measurement is equivalent
to universal quantum computation.
Finally, recent work has considered a constrained
family of Hamiltonians in order to produce new results on allowed
groundstates \cite{stoquastic}.

Label $\mathcal{P}_n$ as the group with elements 
$\{ \pm \sigma_{\otimes J}\}_{J \in \mathcal{J}}$, 
and consider a subgroup $G \subset \mathcal{P}_n$.  
The stabilizer codespace of $G$ is defined \dpo{as}
\begin{equation}
\mathcal{C}(G) \  {\buildrel \mbox{\tiny def} \over =}
\ \bigg\{ \; \ket{\psi} \in \mathcal{H}_n \; | \;
g \ket{\psi} = \ket{\psi} \quad \forall g \in G \; \bigg\} .
\end{equation}
The codespace $\mathcal{C}(G)$ is nonzero \cite{Ike&Mike}
if and only if $G$ is commutative.  Now suppose 
we have a set $S$ of $\pm \sigma_{\otimes J}$ that generate $G$ and
are at most $k$-local.  Given commutativity, we may equally well
think of the codespace as the ground eigenspace of the following
Hamiltonian.
\begin{equation}
\label{eq:Hamiltonian}
H_{\mathcal{C}} \ = \ \sum_{\pm \sigma_{\otimes J} \in S}
\mp \sigma_{\otimes J}
\end{equation}
Even for rather small $k$, in fact even $k=3$, engineering such
a $k$-local Hamiltonian is challenging.  Hence one wishes 
to find a Hamiltonian $H_{\mbox{\footnotesize eff}}$ with the same
groundstate eigenspace as $H_{\mathcal{C}}$ yet which is
$\ell$-local for $\ell<k$.  Such a Hamiltonian 
$H_{\mbox{\footnotesize eff}}$ is called an
\emph{effective Hamiltonian} for $H_{\mathcal{C}}$.
This paper provides conditions under which no 
such effective Hamiltonian exists.

Two applications result.  First, consider that the
Hamiltonian whose groundstate is a stabilizer code has created
an energy-gap to leaving the code.  This energy gap might be
viewed as passive error correction, and our bounds on $\ell$-locality
of effective Hamiltonians become minimum expenses for obtaining
such behavior.  In particular, these results
provide a quantitative argument that
the four-local costs for toric codes \cite{Kitaev} 
and analogous codes for cellulated surfaces
\cite{FreedmanMeyer,bombindelgado} are the best possible.
A second application regards
adiabatic computing, where attempts to drive down the required
$k$-locality of adiabatic algorithms motivates the
search for effective Hamiltonians \cite{gadget}.
In that context, these arguments show that even
effective Hamiltonians must be at least $k$-local for certain
fixed $k$.  However, such bounds are only on effective Hamiltonians
which do not exploit ancillae.  Of course, they still apply to
systems with ancillae if the ancillae are included in a larger
system.

The \dpo{manuscript} is organized as follows.
Two notions of locality of a stabilizer subgroup
of the Pauli group are introduced in \S \ref{sec:exactestimates},
and each notion leads to a theorem constraining the inclusion
of code spaces into the groundstates of Hamiltonians which are
excessively local.  A perturbative variant in \S \ref{sec:perturbation}
shows that if the groundspace of an excessively local Hamiltonian
is too near the stabilizer code, then the gap between the groundstate
eigenvalue and the next distinct eigenvalue is pinched.
Finally, we consider two examples in \S \ref{sec:examples},
namely Calderbank-Shor-Steane codes and stabilizer codes
arising from cellulations of surfaces.

\section{Stabilizer Codes as Exact Groundspaces}
\label{sec:exactestimates}

We write Pauli tensors as $\sigma_{\otimes J} =
\sigma_{j_1} \otimes \sigma_{j_2} \otimes \cdots \otimes \sigma_{j_n}$ 
for $J=j_1 j_2 \ldots j_n$ and each $j_k \in \{0,x,y,z\}$,
where $\sigma_0={\bf 1}$ and the other letters denote the usual
Pauli matrices.  For $\mathcal{J}$ the set of all such indices
$J$, the Pauli group $\mathcal{P}_n$ is $\{ \pm \sigma_{\otimes J} \; | \;
J \in \mathcal{J} \}$.  Thus $|\mathcal{P}_n|=2 \cdot 4^n$,
every nonidentity element $g \in \mathcal{P}_n$ has $g^2 = {\bf 1}$,
and all elements of $\mathcal{P}_n$ commute or anticommute.
For $G$ a subgroup of $\mathcal{P}_n$, the stabilizer code
of $G$ is the subspace of $\mathcal{H}_n$ which is the intersection
of the $+1$ eigenspaces of all $g \in G$.  It is known that
the code space, say $\mathcal{C}$, is nonzero if and only if
$G$ is commutative \cite{Ike&Mike}.  It is common to 
refer to $G$ as a stabilizer group of $\mathcal{C}$ when
(conversely) $\mathcal{C}$ is the intersection of the $+1$
eigenspaces of $g \in G$.  Being less precise, a commutative
subgroup $G \subseteq \mathcal{P}_n$ is a stabilizer group
(of some nonzero $\mathcal{C}$).

The discussion requires additional background on
stabilizer codes.  In particular, we highlight
the following facts.

\medskip

\noindent
{\bf Lemma:} {\bf [See \cite[\S10.5.1]{Ike&Mike}.]}
\emph{
(i) Let $G \subseteq \mathcal{P}_n$ and
$\Pi_G \ = \ (1/|G|) \sum_{g \in G} \; g$.
Then for commutative $G$, $\Pi_G$ is a projector onto
the code space of $G$.  Else $\Pi_G = {\bf 0}$.
(ii)  If $\sigma$ and $-\sigma$ are both in $G$, 
then the code space is trivial.
}

\medskip

\noindent
{\bf Proof:}
The first item is proven in
the citation.  For the second, the hypothesis requires
$-{\bf 1}=(\sigma)(-\sigma) \in G$.  Thus
$\mbox{Trace}(\Pi)=\mbox{Trace}({\bf 1} - {\bf 1}) =0$, since
every element of $\mathcal{P}_n$ other than $\pm {\bf 1}$
is traceless.  Since the projector $\Pi$ is traceless,
it is zero.  Hence its target, the code space, is trivial.
\hfill $\Box$

Use $\mbox{wt}(g)$ for $g \in \mathcal{P}_n$ to denote the
number of $\sigma_x$, $\sigma_y$, and $\sigma_z$ factors of the
tensor product.  In particular, 
$\mbox{wt}(-{\bf 1})=0$.  Also,
$\mbox{wt}(g_1 g_2) \leq \mbox{wt}(g_1)+\mbox{wt}(g_2)$,
since any qubit whose tensor factors are ${\bf 1}$ in $g_1$ and
$g_2$ will have tensor factor ${\bf 1}$ in their product.
Finally, for $S \subset \mathcal{P}_n$, we use
$\langle S \rangle$ to denote the subgroup generated by the elements
of $S$.  This is standard notation from abstract algebra, and we hope
that context will make clear that it is not the Dirac notation
for the expectation of an operator $S$.

Recall from Equation \ref{eq:Rbbcombo} that any
Hamiltonian on $n$ qubits may be written as a real linear
combination of Pauli tensors.  The Hamiltonian is $k$-local
if the degree of no monomial summand exceeds $k$.  This is a
measure of complexity of the Hamiltonian and physical systems
that realize it, in that $k$-local Hamiltonians require at most
$k$-qubits to interact during any infinitesimal time.

This section presents two results which
argue that Hamiltonians
whose groundstate captures a stabilizer code must be at least as complicated
as the underlying stabilizer group.
The complication of Hamiltonians is measured in $k$-locality.
On the other hand, two reasonable
definitions of the $k$-locality of stabilizer 
group are considered in separate subsections.
These two measures are motivated by earlier work
\cite{graph} and so are denoted
$\delta(G)$, a lower bound on the weight of $g \in G$,
and $\eta(G)$, in principle an upper bound.
We begin with $\delta(G)$.

\subsection{Lower Bound Case}
\label{subsec:lower}

\medskip

\dpo{We now define a quantity
$\delta(G)$ that may be viewed as a lower bound on the $k$-locality
of a stabilizer group. } 

\medskip

\noindent
{\bf Definition:} Let $G \subseteq \mathcal{P}_n$ be a subgroup.  
Then $\delta(G) = \mbox{min }\{ \mbox{wt}(g) \; | \;
g \in G, g \neq {\bf 1} \}$.

\medskip

The next result implies 
that any Hamiltonian
$H$ whose groundstate is the stabilizer code must be at least
$\delta(G)$ local.  To see this, normalize so that $H$ is
traceless by subtracting the appropriate multiple of ${\bf 1}$.
The groundspace of the traceless Hamiltonian is then a negative
eigenspace.

\medskip

\vbox{
\noindent
{\bf Theorem 1:}  
\emph{
Let $G$ be a stabilizer group and let
$H$ be a traceless Hamiltonian on $n$-qubits which
is $k$-local for $k < \delta(G)$.
Let $\mathcal{V}_- \subset \mathcal{H}_n$ be the direct sum of
eigenspaces of $H$ corresponding to negative eigenvalues.  
Then the codespace of $G$ is not contained within $\mathcal{V}_-$.
}
}

\medskip

\noindent
{\bf Proof:}  Let $\{ \ket{\psi_j}\}_{j=1}^L$
form a basis for the codespace.  Recall $\Pi$
from the proof of the \dpo{Lemma:}
\[
\Pi \ = \sum_{j=1}^L \ket{\psi_j}\bra{\psi_j} \ = \ 
\ (1/|G|) \sum_{g \in G} \; g \, .
\]
While the second expression is an orthogonal decomposition
of a projector, the third is a well known formula
for a projector onto the code space
\cite[\S10.5.1]{Ike&Mike}.

Recall the decomposition of the Hamiltonian $H$ according
to Equation \ref{eq:sumofpaulis} in the introduction.
\[
H \ = \ \sum_{J \in \mathcal{J}} t_J \sigma_{\otimes J},
\quad t_J \in \mathbb{R}.
\]
The traceless condition forces $t_{00 \ldots 0}=0$,
since for $J \neq 00\ldots0$ we have $\mbox{Trace}(\sigma_{\otimes J})
= \prod_{k=1}^n \mbox{Trace}(\sigma_{j_k}) =0$.
If some coefficient $t_J$ is nonzero, then by hypothesis
$\#J \leq k < \delta(G)$.

The estimate follows by considering 
$\mbox{Trace}(\sigma_{\otimes J} g)$ for $g \in G$ and $t_J \neq 0$.
Then $\sigma_{\otimes J} g  \neq \pm {\bf 1} \in \mathcal{P}_n$, 
since the product has weight at least one.  
For $g$ has weight 
at least $\delta(G)$ while $\sigma_{\otimes J}$ has weight at most
$k < \delta(G)$, and $\sigma_{\otimes J}=\sigma_{\otimes J}^{-1}$,
hence $\mbox{wt}(g)=\mbox{wt}(\sigma_{\otimes J}^{-1}
\sigma_{\otimes J} g ) \leq \mbox{wt}(\sigma_{\otimes J} g)
+\mbox{wt}(\sigma_{\otimes J})$ or $\mbox{wt}(\sigma_{\otimes J} g) \geq  
\delta(G) - k$.  Therefore $\mbox{Trace}(\sigma_{\otimes J} g)=0$
since $\mbox{Trace}(h)=0$ for any $h \in \mathcal{P}_n-\{ \pm{\bf 1}\}$.
The right hand equality of the
equation below follows.
\begin{equation}
\label{eq:basis}
\sum_{j=1}^L \langle \psi_j | H | \psi_j \rangle \ = \ 
\mbox{Trace} \big( \; \Pi H \; \big) \ = \ 
(1/|G|) \sum_{g \in G} \mbox{Trace} \big( \; g H \; \big) \ = \ 0 \, .
\end{equation}
Now if $\{ \ket{\psi_j}\}_{j=1}^L \subseteq \mathcal{V}_-$,
then each term at the far left of Equation \ref{eq:basis} would 
be negative, leading to a contradiction.
\hfill $\Box$

\subsubsection*{How might one compute $\delta(G)$?}

\dpo{We now} sketch how one might compute $\delta(G)$,
using the stabilizer 
check matrix $A$ of the stabilizer code.
Thus $A=(A_X | A_Z) \in (\mathbb{F}_2)^{m \times n}$
corresponding to the choice of generators $\{g_j\}_{j=1}^m$,
i.e. $G=\langle \{g_j\}_{j=1}^m \rangle$.
A $1$ in row $k$ of column $j$ of $A_X$ corresponds
to a factor of $\sigma_x$ in qubit position $k$ of generator $g_j$,
and $A_Z$ is similar.
(See \cite[eqn. (10.112)]{Ike&Mike} or \cite[\S2.2.3]{biggraph}.)
Since $m$ is the number of generators for $G$ and 
$g^2 = {\bf 1}$ for
any $g \in \mathcal{P}_n$,
one way to calculate $\delta(G)$ would be to enumerate all
$2^m$ products of generators.  A possible optimization
of this approach would be to delete generators until the set 
$\{g_j\}_{j=1}^m$ is \emph{minimal}, i.e. until the number of rows 
of $A$ is also its rank.

We present a different approach.
Namely, suppose that a $p$-local tensor product $g=\pm \sigma_{\otimes J}$ 
is in $G$.  
The support of $g$ will be given by
$\mbox{supp}(g)=\{ k \in \{1,2,\ldots, n\} \; | \; j_k \neq 0 \}$,
so that $|\mbox{supp}(g)|=p$.  Then for $v$ an indicator
vector of which generators occur in the product for $g$,
$v^T A = (w_X | w_Z)$ has $w_X$ and $w_Z$ zero outside entries
indexed by $S$.  Now label $A_S$ as that matrix with the columns
of $A_X$ and $A_Z$ corresponding to $S$ replaced by zero entries.
Then $v^T$ is a left-null vector of $A_S$ but not of $A$.
On the other hand, any left-null vector of $A$, say
$w$ with $w^T A =0$, must also satisfy $w^T A_S = 0$.
Thus $\mbox{rank}(A_S) < \mbox{rank}(A)$.

\medskip

\noindent
{\bf Example:}  Consider $G=\langle X \otimes I \otimes Z,
I \otimes Z \otimes X \rangle$, for which $\delta(G)=2$.
Taking a basis for $A$ according to the generating set above yields
the \dpo{equation}
\begin{equation}
A \ = \ 
\left(
\begin{array}{rrrrrr}
1 & 0 & 0 & 0 & 0 & 1 \\
0 & 0 & 1 & 0 & 1 & 0 \\
\end{array}
\right) .
\end{equation}
Say $S=\{1,3\}$, since the first row of $A$ recovers
the two-local $X \otimes I \otimes Z$ supported on
these qubits.  \dpo{Then}
\begin{equation}
A_S \ = \ 
\left(
\begin{array}{rrrrrr}
0 & 0 & 0 & 0 & 0 & 0 \\
0 & 0 & 0 & 0 & 1 & 0 \\
\end{array}
\right).
\end{equation}
Thus the existence of this two-local element of $G$
has caused $\mbox{rank}(A_S) < \mbox{rank}(A)$, which might
also be inferred due to the left-null vector $v^T = (1 0)$.

\vbox{
\vspace{.1in}
\noindent
\hrulefill

\noindent
{\bf Algorithm:}  Computing $\delta(G)$

\noindent
\hrulefill

\noindent
For $k=1,\dots,m$\newline
\hbox{\ \  } For each $S\subseteq\{1,\dots,n\}$ with $\# S = k$ do: \newline
\hbox{\ \ \ \ }
Compute $A_S$ by deleting columns of $A$ corresponding to $S$.
\newline
\hbox{\ \ \ \ }
 If $\mbox{rank}(A_S) < \mbox{rank}(A)$ then \newline
\hbox{\ \ \ \ \ \ } Return $\delta(G)=k$ and exit. \newline
\hbox{\ \ } End for. \newline
End for.  \newline
\noindent
\hrulefill
}

The algorithm above for computing $\delta(G)$
is polynomial in $n$, at least
if $\delta(G) \in O(1)$.  (Cf. \cite{graph}.)  
Note however that it is not a polynomial time algorithm
should $\delta(G) \in \Omega(n)$, since then the loop will
loop over (a nonnegligible fraction of) the power set of
$\{1,2,\ldots,n\}$.

\subsection{Upper Bound Case}
\label{subsec:upper}

This section considers $\eta(G)$, which is an upper bound on the 
$k$-locality of $G$.  However,
we do not define $\eta(G)$ to be
the maximum of weights of $g \in G$.  For
\hbox{$\ket{+}^{\otimes n} = [2^{-1/2} ( \ket{0}+\ket{1})]^{\otimes n}$}
spans the one-dimensional code space of
the stabilizer group generated by 
the $n$ Hermitian Pauli tensors
$(\sigma_x)_j={\bf 1} \otimes {\bf 1} \otimes \cdots \otimes
\sigma_x \otimes \cdots \otimes {\bf 1}$
with a single Pauli-$X$ on qubit $j$.
Then $G$ contains $\sigma_x^{\otimes n}$ of
weight $n$, yet $\ket{+}^{\otimes n}$ is local.  Thus 
to get a useful definition of an upper bound we
resort to a minimax construction, taking the minimum over all
generating sets of $G$ of the maximum $k$-locality in a given set.
The following definition (Cf. \cite{graph}) is equivalent to 
that minimax.

\medskip

\noindent
{\bf Definition:}  For $S \subseteq \mathcal{P}_n$,
let $\langle S \rangle$ denote the subgroup generated by $S$.
Let $G \subseteq \mathcal{P}_n$ be a stabilizer
group with nontrivial codespace.  Then $\eta(G)$ is the
minimal $\nu$ such that
\hbox{$\langle \{ g \in G \; | \; \mbox{wt}(g) \leq \nu \} \rangle = G$}.

\medskip

\dpo{Next we define a subgroup $G(\overline{b})$ and related notation.}

\noindent
{\bf Definition of $G(\overline{b})$:}  
Let $G\subseteq \mathcal{P}_n$ be a commutative
subgroup and $\nu < \eta(G)$.  We label
\hbox{$G_\nu = \langle \{ g \in G \; | \; \mbox{wt}(g) \leq \nu \} \rangle$}.
Fix a minimal generating set so that (i)
$G_\nu = \langle \{g_j\}_{j=1}^s \rangle$
and (ii) $\mbox{wt}(g_j)\leq \nu$ for each $j$.
Extend this to a minimal generating set so that
$G=\langle \{ g_j \}_{j=1}^t \rangle$, where the 
$g_j=\pm \sigma_{\otimes J}$ may be Pauli tensors of
any degree if $j>s$.
For a bitstring $\overline{b} \in \mathbb{F}_2^{n-s}$,
we label a new subgroup of $\mathcal{P}_n$:
\begin{equation}
G(\overline{b}) \ = \ \langle
\{ g_1,g_2,\ldots,g_s,(-1)^{b_{s+1}}g_{s+1},\ldots,(-1)^{b_t}g_t\}
\rangle .
\end{equation}
The generating set for $G(\overline{b})$
above is also minimal \cite{Ike&Mike}.
The dependence of $G(\overline{b})$ on $\eta$ and
on the (ordered) sequence of generators $\{g_j\}_{j=1}^t$
will be left implicit.

Say $\nu < \eta(G)$.  We do not have a result
which prohibits certain groundstate eigenspaces for
$\nu$-local Hamiltonians as would be the case  
if $\nu<\delta(G)$.  Yet there is a result which is similar to
this, namely that the eigenspaces of $\nu$-local Hamiltonians
may not distinguish $G$ and other extensions $G(\overline{b})$.
This might be of independent interest and will also imply
a result similar to the previous one, except that
the relevant energies of the traceless effective Hamiltonian
must be positive rather than merely nonnegative.

\medskip

\vbox{
\noindent
{\bf Theorem 2:}
\emph{Let $\nu < \eta(G)$.
Let $\Pi_G$ and $\Pi_{G(\overline{b})}$ be projectors on the respective
codespaces, where $b$ is a bitstring and $G(\overline{b})$
is defined above.
Then for any traceless $\nu$-local Hamiltonian $H$\dpo{,}
\begin{equation}
\label{eq:blind}
\mbox{Trace} \big( \; \Pi_G H \; \big) \ = \ 
\mbox{Trace} \big( \; \Pi_{G(\overline{b})} H \; \big) .
\end{equation}
}
}

\medskip

\noindent
{\bf Proof:}
Let $\sigma \in \mathcal{P}_n$ such that $\mbox{wt}(\sigma) \leq \nu$.
It suffices to show that
$\mbox{Trace} \big( \; \Pi_G \sigma \; \big)  =  
\mbox{Trace} \big( \; \Pi_{G(\overline{b})} \sigma \; \big)$.
We prove this formula using a case study.

\smallskip

\noindent
{\bf Case 1:}  Suppose either $\sigma \in G$
or $-\sigma \in G$ or both.  
Each such element is in $G_\nu$ due to its
weight, hence each such element is also an element
of $G(\overline{b})$.

Since by hypothesis $G$ has a nontrivial codespace,
both $\sigma$ and $-\sigma$ are not in $G$.  (Else ${-\bf 1}
\in G$ and $\mbox{Trace}(\Pi_G)={\bf 0}$ contradiction.)
As a remark,  $\mbox{Trace} \big( \; \Pi_G \sigma \; \big)  =  
\mbox{Trace} \big( \; \Pi_{G(\overline{b})} \sigma \; \big)$
nonetheless holds in this subcase as $0=0$.

Thus say $\sigma \in G$ with $-\sigma \not\in G$ or
vice-versa.  Then each trace is $\pm 2^{t-n}$, since 
(i) $\mbox{Trace}(g_1 g_2)=0$
whenever $g_1,g_2 \in \mathcal{P}_n$ and $g_1 \not\in \{g_2,-g_2\}$
and (ii) the size of the minimal generating sets demand
$2^{n-t}=\# G = \# G(\overline{b})$ \cite{Ike&Mike}.

\smallskip

\noindent
{\bf Case 2:}  Suppose $\sigma \not\in G$ and $-\sigma
\not\in G$.  Then 
$\mbox{Trace} \big( \; \Pi_G \sigma \; \big) = 0$.
It would suffice to show that
$\sigma \not\in G(\overline{b})$ and $-\sigma \not\in G(\overline{b})$.

Assume by way of contradiction
that $\sigma \in G(\overline{b})$.
Then for a bit-string $\overline{c}=c_{s+1}c_{s+2}\ldots c_t$,
we \dpo{have}
\begin{equation}
\sigma \ = \ \prod_{j=1}^s g_j^s \; \prod_{j=s+1}^t
\bigg( \; (-1)^{b_j} g_j \; \bigg)^{c_j} .
\end{equation}
Since $\prod_{j=s+1}^t (-1)^{b_j c_j} \in \{1,-1\}$,
either $\sigma \in G$ or else $-\sigma \in G$.  Contradiction.
The case that $-\sigma \in G(\overline{b})$ is similar.
\hfill $\Box$

\medskip

\vbox{
\noindent
{\bf Corollary 3:}  
\emph{Suppose $G$ and $G(\overline{b})$ as in the Theorem.
Suppose that $H$ is a
traceless, $\nu$-local Hamiltonian
for $\nu<\eta(G)$.  Partition
\hbox{$\mathcal{H}_n=\mathcal{V}_- \oplus \mathcal{V}_0 \oplus \mathcal{V}_+$}
into positive, zero, and negative eigenspaces of $H$.
\begin{itemize}
\item  If the codespace of $G$ is contained within 
$\mathcal{V}_-\oplus \mathcal{V}_0$,
then the codespace of $G(\overline{b})$ is contained within
$\mathcal{V}_-\oplus \mathcal{V}_0$.
\item  Let $\mathcal{C}(\overline{b})$ denote the codespace of
$G(\overline{b})$.
If $\cup_{\overline{b}} \mathcal{C}(\overline{b})$ spans
$\mathcal{H}_n$, then the codespace of $G$ is not a ground
eigenspace of any $k$-local Hamiltonian $H$.
\end{itemize}
}
}

\subsection*{How might one compute $\eta(G)$?}

Recall the earlier algorithm to compute
$\delta(G)$ using $A=(A_X | A_Z) \in \mathbb{F}_2^{m \times n}$.
This section produces a similar algorithm for $\eta(G)$
using linear algebra.  However, we first need some more notation.
Namely, although the subset of $k$-local elements within
$G$ do not form a subgroup, those elements which only
affect any collection of $k$-qubits do.  The algorithm for
$\eta(G)$ represents these subgroups 
as matrices and then uses algebra to decide whether their
union generates $G$.

\medskip

\noindent
{\bf Definition:}  Recall the notation 
$\sigma_{\otimes J}=\sigma_{j_1} \otimes \cdots \otimes \sigma_{j_n}$
for $J=j_1j_2\ldots j_n$ and $j_k \in \{0,x,y,z\}$,
where $\sigma_0={\bf 1}$ and the other sigmas denote the
appropriate Pauli matrices. 
The \emph{support of $\pm \sigma_{\otimes J}$}, 
$\mbox{supp}(\pm \sigma_{\otimes J})$,
is $S = \{ k \; | \; j_k \neq 0 \} \subseteq \{1,2,\ldots,n\}$.
Label the subgroup $\mathcal{P}_S = \{ g \in \mathcal{P}_n \; | \;
\mbox{supp}(g) \subseteq S \}$.
Also set $G_S = G \cap \mathcal{P}_S$.

\medskip

Henceforth, suppose $G$ is fixed with \emph{nontrivial codespace},
so that by the Lemma $g \in G$ demands $-g \not\in G$.
This creates a map from the row space of $A$ to $G$.
Indeed, since rows of $A$ represent generators of $G$,
the fact that the row vector
$(b_1b_2 \cdots b_n  c_1c_2\ldots c_n)$
lies within the row space implies
$\pm (\sigma_x^{b_1} \otimes \sigma_x^{b_2} \otimes \cdots 
\otimes \sigma_x^{b_n})
(\sigma_z^{c_1} \otimes \sigma_z^{c_2} \otimes \cdots 
\otimes \sigma_z^{c_n})$
is an element of $G$.  Furthermore, although the $2n$ bitstring
does not make clear the choice of sign, the Lemma asserts that
it is unique.  Now recall that $A_S$ is the matrix $A$ except
that columns corresponding to $S \subseteq \{1,2,\ldots,n\}$
have been replaced with zero columns.  As a consequence
of the unique sign choice of the Lemma, any $2n$-bit string
in the rowspace supported on positions corresponding to
$S$ likewise determines an element of $G_S$.  For $\overline{S}$
the complement of $A$, such an element
of the rowspace might be constructed by creating a left-null
vector of $A_{\overline{S}}$.

\medskip

\vbox{
\noindent
\hrulefill

\noindent
{\bf Algorithm:}  Computing $\eta(G)$.

\noindent
\hrulefill

\noindent
for $k=1:n$ \newline
\hbox{\ \  } $A_k\in \mathbb{F}_2^{0\times 2n};$ \newline
\hbox{\ \  } for $S \subseteq \{1,2,\ldots,n\}$ with $|S|=k$ \newline
\hbox{\ \ \ \ }Compute $N_S$, a matrix whose rows span the
left-null space of $A$. \newline
\hbox{\ \ \ \ }Compute $B_S=N_S A_S$, 
the matrix encoding $G_S$.  \newline
\hbox{\ \ \ \ }Set $A_k = \left( \begin{array}{c} A_k \\ B_S \\ 
\end{array}\right)$.  \newline
\hbox{\ \ \ \ }if $\mbox{rank}(A_k) = \mbox{rank}(A)$ \newline
\hbox{\ \ \ \ \ \ }return $\eta(G)=k$  \newline
\hbox{\ \ \ \ }End if.  \newline
\hbox{\ \ }End for.  \newline
End for.  \newline

\noindent
\hrulefill
}

\medskip

\section{Gap-Pinching when Approximating Stabilizer Codes}
\label{sec:perturbation}

Our two earlier results limit those cases in which a
stabilizer code lies within the groundstate of
a Hamiltonian whose $k$-locality is less than
some measure of the locality of the stabilizer group.
The two measures of the group's locality
were $\delta(G)$ and $\eta(G)$,
where $\delta(G) \leq \eta(G)$.  The result for $\delta(G)$
is stronger even though it applies to fewer Hamiltonians,
in that it prohibits stabilizer codes within groundstates.  In contrast,
the more widely applicable result regarding $\eta(G)$ 
allows zero eigenspaces of traceless $H$ to contribute
and also requires some study of auxilliary stabilizer groups
$G(\overline{b})$.  Neither of these results was perturbative.
We next present a result which limits those cases in which
a stabilizer code is merely close to the groundstate of
a Hamiltonian which is more local than the code.  More precisely,
we argue that the groundstate eigenspace of such a Hamiltonian
lacks stability, in that the gap between the lowest two
distinct eigenvalues is small when compared to the total energy
of the system.  Similar results regarding stabilizer codes
for graph states are known \cite{graph}.

The notation below will be fixed while discussing
the perturbative result.
\begin{itemize}
\item  Let $q=\mbox{dim}_{\mathbb{C}} \mathcal{C}(G)$,
with $\mathcal{C}(G)$ the
stabilizer code of $G$.  For a graph state, $q=1$.
\item  Defining $\nu < \eta(G)$  and $G_\nu$ as above,  we
let $r=\mbox{dim}_{\mathbb{C}} \mathcal{C}_\nu$ where 
$\mathcal{C}_\nu = \mathcal{C}(G_\nu)$.  Similarly
let $\mathcal{C}=\mathcal{C}(G)$.  
Since $G_\nu \subseteq G$, also 
$\mathcal{C}_\nu \supseteq \mathcal{C}$
and thus $r \geq q$.
\item  Consider $H$ a $\nu$-local
Hamiltonian with $\Pi_H$ the projection
onto its groundstate eigenspace.
\item  $\Pi_G$ and $\Pi_{G_\nu}$ are projections onto the appropriate 
stabilizer codespaces.
\item  The trace norm $\| \ast \|_{\mbox{\footnotesize tr}}$ on Hermitian 
matrices
is that norm induced by the inner product 
$H_1 \bullet H_2 = \mbox{Trace}(H_1 H_2^\dagger)
= \mbox{Trace}(H_1 H_2)$.
\end{itemize}
In addition to the setup, we should note that one way to quantify the 
distance between
the code of $G$ and the groundstate of $H$ is to compute the trace norm of the
difference of the projectors onto each space.

\medskip

\noindent
{\bf Theorem:}
\emph{
Let $G \subseteq \mathcal{P}_n$ have a code space of
dimension} $q>0$.  Let $\nu < \eta(G)$.
Then any traceless $\nu$-local Hamiltonian $H$ 
whose groundstate eigenspace is $q$-dimensional 
satisfies the following inequality on the trace norm distance
between the projectors $\Pi_G$ and $\Pi_H$ onto the codespace of
$G$ and the groundstate eigenspace of $H$ respectively.
\begin{equation}
\| \; \Pi_G - \Pi_H \; \|_{\footnotesize \rm tr}  \ \geq \ 
\frac{q}{\| \vec{E} \|_2} 
\bigg(\; \frac{E_0+E_1+E_2+\cdots +E_{r-1}}{r} - E_0 \; \bigg)
.
\end{equation}
Here, $E_0 \leq E_1 \leq \cdots
\leq E_{2^n-1}$ is the eigenspectrum of $H$ (with multiplicity)
and $\| \vec{E} \|_2 = \mbox{Trace}(H^2)^{1/2}
= ( E_0^2 + E_1^2 + \cdots + E_{2^n-1}^2)^{1/2}$.  Also,
$r$ denotes the dimension of the codespace of the group
$G_\nu \subseteq G$ generated by $\nu$-local elements.

\medskip

\noindent
{\bf Proof:}  The first step is to check that due to the locality 
condition on $H$, we have
$\mbox{Trace}( \Pi_G H) = (q/r) \mbox{Trace}( \Pi_{G_\nu} H)$.  Since 
all elements
of $G$ and $G_\nu$ that are at most $\nu$-local
coincide, the following traces are equal.
\begin{equation}
\mbox{Trace} \bigg( \; H \sum_{g \in G} g \; \bigg) \ = \ 
\mbox{Trace} \bigg( \; H \sum_{g \in G_\nu} g \; \bigg).
\end{equation}
The projectors should be normalized by $\# G$ and $\# G_\nu$ respectively.
If $m$ is the number of rows of a stabilizer check 
matrix for $G$ arising
from a minimal generating set and $m_\nu$ is similar for $G_\nu$,
then $\# G = 2^m$ and $\# G_\nu=2^{m_\nu}$ \cite{Ike&Mike}.  Furthermore
$q=2^n/2^{m}$ and $r=2^n/2^{m_\nu}$.  Thus appropriately
normalizing the above equation produces the desired equality.

Now let $\Pi$ be any projection onto an $r$-dimensional space.  Since
$E_0 \leq E_1 \leq \cdots \leq E_{r-1}$ are the $r$ least eigenvalues of $H$, 
we have
the \dpo{inequality}
\begin{equation}
\mbox{Trace}( \Pi H) \ \geq \ (E_0 + E_1 + E_2 + \cdots + E_{r-1} ) .
\end{equation}

Recall that the inner product associated to the trace norm has a 
Schwarz inequality.
This is the final fact required for the following sequence of inequalities.
\begin{equation}
\begin{array}{lcl}
\| \vec{E} \|_2 \| \Pi_G - \Pi_H \|_{\mbox{\footnotesize tr}}
& = & \| H \|_{\mbox{\footnotesize tr}}
\| \Pi_G - \Pi_H \|_{\mbox{\footnotesize tr}} \\
& \geq & \mbox{Trace} \big( \; ( \Pi_G - \Pi_H )  H \; \big) \\
& = & (q/r) \mbox{Trace}( \Pi_{G_\nu} H) - \mbox{Trace}(\Pi_H H) \\
& \geq & (q/r) ( E_0 + E_1 + E_2 + \cdots + E_{r-1} ) - q  E_0 .\\
\end{array}
\end{equation}
Appropriate manipulations of the inequality between the first and last
expression of the sequence above produces the result.  \hfill $\Box$

\noindent
{\bf Corollary:}
\emph{
Let the traceless Hamiltonian $H$ and code $G$ satisfy all hypotheses
of the theorem, including excessive locality of $H$ as compared
to $\eta(G)$.
Label the spectral gap of $H$ as
$\Delta E = E_q-E_0$, recalling $E_0=E_1=E_2=\cdots=E_{q-1}$.
The following estimate holds:
\begin{equation}
\|\; \Pi_G - \Pi_H \; \|_{\footnotesize \rm tr}  \ \ \geq \ \ 
q \; \| \vec{E} \|_2^{-2} \; \big( (r-q)/r \big) \; \; \Delta E .
\end{equation}
In particular, if $\epsilon > \|\; \Pi_G - \Pi_H \; \|_{\footnotesize \rm tr}$ 
and $r$ and $q$
are treated as constants, then the gap is pinched in the sense that
$\Delta E \in O( \epsilon \| \vec{E}\|_2^2)$.
}

\noindent
{\bf Proof:}
Notice that for $j \geq q$, $E_j \geq E_0 + \Delta E$.  
The term inside
the parentheses of the Theorem is bounded below by a multiple of this gap
(Cf. \cite{graph})\dpo{:}
\begin{equation}
\begin{array}{lcl}
\left( \frac{ E_0 + E_1 + E_2 + \cdots + E_{r-1}}{r} \right) - E_0 & \geq 
&  \big( (q/r)E_0 + ((r-q)/r) E_q \big) - E_0 \\
& = & \big((q-r)/r\big) E_0 + \big((r-q)/r\big) E_q \\
& = & \big((r-q)/r\big) \Delta E . \\
\end{array}
\end{equation}
\hfill $\Box$

The pinching bound of the Corollary is weak
in the following sense.  (Cf. \cite{graph}.)  
Effective Hamiltonians are used to
approximate lower energy eigenstates while ignoring
higher energy eigenstates.  Thus the large total
energy $\| \vec{E} \|_2^2$ is not a concern.
On the other hand, the Corollary
also argues that the higher energy eigenstates can
not be (entirely) irrelevant to such approximations.

\section{Examples}
\label{sec:examples}

This section considers the computation of the quantities
$\delta(G)$ and $\eta(G)$ used in the effective Hamiltonian bounds
in special cases.  We first note simplifications for a broad class of
codes that includes CSS codes and also topological
orders on surfaces.  The topological order case requires
further attention, in that answers should be computable using
only the cellulation of the surface.  
The codes depend on the cellulation rather than the the topology
(i.e. genus) of the surface, and the same is true of 
$\delta(G)$ and $\eta(G)$.

\begin{figure}
\begin{center}
\includegraphics[scale=0.5]{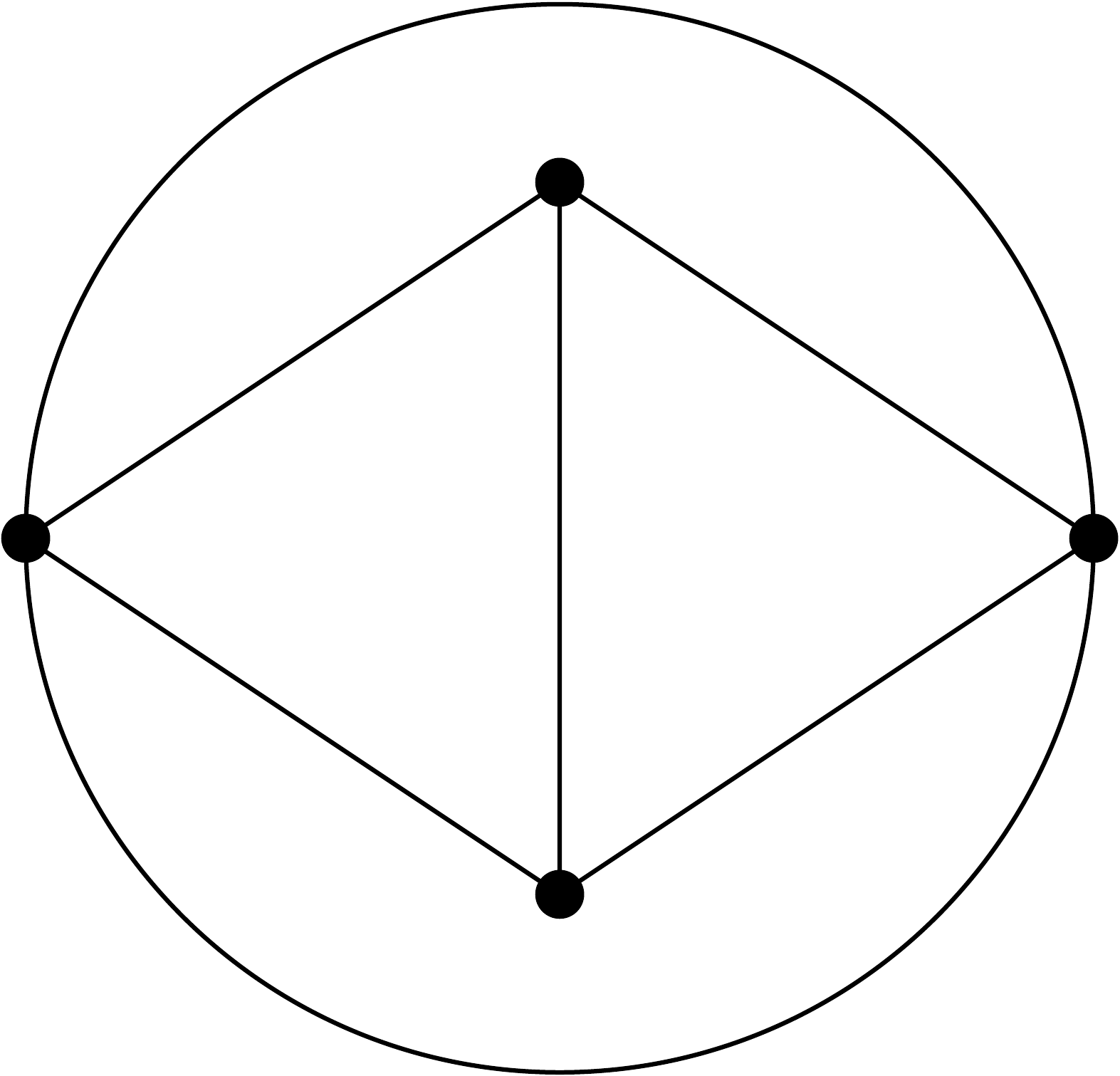}
\end{center}
\caption{\label{fig:1} 
\emph{
A counterexample to the conjecture that $\delta(G)$ is the
mininum of the valences of the one-skeleta of the cellulation
$\Gamma$ and the dual cellulation $\Gamma^\ast$ results as follows.
Cellulate a disc as above.  Cellulate a sphere with the top and
bottom each copies of this disc.  Then $\delta(G)=2$, where
the minimal boundary is the two-edge circle which bounds either disc.
Yet the minimum of the valences is three.
}
}
\end{figure}

\subsection{Calderbank-Shor-Steane codes}

\vbox{
\noindent
{\bf Definition:}  Let $\mathcal{P}_{X,n}
= \langle \{ \sigma_{x,j} \}_{j=1}^n \cup \{ -\sigma_{x,j} \}_{j=1}^n
\rangle$ be the subgroup
of $\mathcal{P}_n$ containing Pauli tensors with only Pauli $X$
factors, and let $\mathcal{P}_{Z,n}$ be similar.
A stabilizer group $G$ is $XZ$ split if 
$G=\langle \{ g_j \}_{j=1}^m \rangle$ where for each $j$ either
$g_j \in \mathcal{P}_{X,n}$ or $g_j \in \mathcal{P}_{Z,n}$.
Perhaps upon reordering, this produces a block-diagonal 
stabilizer check
matrix with blocks $A_X$ and $A_Z$ defined by the following equation:
\begin{equation}
A \ = \ \left(
\begin{array}{cc}
A_X & {\bf 0} \\
{\bf 0} & A_Z \\
\end{array}
\right) .
\end{equation}
We also label $G_X = \mathcal{P}_{X,n} \cap G$ and
$G_Z = \mathcal{P}_{Z,n} \cap G$.
}

All Calderbank-Shor-Steane codes \cite[\S10.4.2]{Ike&Mike}
\cite{CalShor,Steane} are $XZ$-split.
Indeed, suppose $\mbox{CSS}(C_1,C_2)$ is the code arising from
classical codes $C_1$ and $C_2$, where $C_1$ corrects bit-flips
and $C_2$ phase-flips.  Then for $A_1$ the parity check matrix
of $C_1$ and $A_2$ the parity check matrix of the dual code
$C_2^\perp$, we have a stabilizer check matrix
$A=\mbox{diag}(A_1,A_2)$ for  $\mbox{CSS}(C_1,C_2)$.
The converse only holds in a technical sense\footnote{
Any $XZ$-split code might be associated to $\mbox{CSS}(C_1,C_2)$
for some classical codes $C_1$ and $C_2$, yet the ratio of logical
to encoding bits of these classical codes would be arbitrary.
Thus we retain $XZ$-split as a separate concept.}.
Next, we study $\delta(G)$ and $\eta(G)$ for $XZ$-split codes.

\medskip

\vbox{
\noindent
{\bf Proposition:}  
\emph{
Suppose that $G$ has nonzero
code space. Label 
\hbox{$G_X G_Z = \{ g_x g_z \; | \; g_x \in G_X, g_z \in G_Z\}$}.
\begin{itemize}
\item ($G$ is $XZ$-split) $\Longleftrightarrow$ ($G=G_X G_Z$).
\item If $G$ is $XZ$ split, then
$\delta(G) = \mbox{min }\{ \delta(G_X), \delta(G_Z) \}$.
\item If $G$ is $XZ$ split, then
$\eta(G) = \mbox{max }\{ \eta(G_X), \eta(G_Z) \}$.
\end{itemize}
}
}

\medskip

\noindent
{\bf Proof:}
For the first item, since $G_X \subseteq G$ and $G_Z \subseteq G$,
we must have $G_X G_Z = \{ g_x g_z \; ;\; g_x \in G_X, g_z \in G_Z\}$
within $G$.  For the opposite containment, the generators guaranteed
by the $XZ$-split condition show that $G=\langle G_X G_Z \rangle$.
On the other hand, finite products
of elements in $G_X G_Z$ lie in $G_X G_Z$, since $G$ is commutative.

For the second item, the minimum is greater than
$\delta(G)$ since $G_{X} \subseteq G$ and
$G_{Z} \subseteq G$ imply
$\delta(G_X) \leq \delta(G)$ and
$\delta(G_Z) \leq \delta(G)$.  On the other hand,
let $g \in G$.  Then $g=g_x g_z$ and
$\mbox{wt}(g) \geq \mbox{max}\{\mbox{wt}(g_x),\mbox{wt}(g_z)\}$
since any qubit on which either $g_x$ or $g_z$ has a nontrivial
tensor factor will have a nontrivial factor in the product for
$g$.

For the last item, let $G_X=\langle \{ g_{x,j} \}_{j=1}^{m_x} \rangle$ and 
$G_Z=\langle \{ g_{z,j} \}_{j=1}^{m_z} \rangle$ be generating sets
chosen to be at most $\eta(G_X)$ local and $\eta(G_Z)$ local.
Since $G=G_X G_Z$, we have
$G=\langle\{ g_{x,j} \}_{j=1}^{m_x} \cup \{ g_{z,j} \}_{j=1}^{m_z}\rangle$.
Thus $\eta(G) \leq \mbox{max}\{\eta(G_x),\eta(G_z)\}$.

On the other hand, assume by way of contradiction that
$G=\langle \{ g_j \}_{j=1}^m \rangle$ where every $g_j$ has weight 
strictly less than
$\mbox{max }\{ \eta(G_X), \eta(G_Z) \}$.  Writing $g_j = g_{j,x} g_{j,z}$
produces generating sets $G_X = \langle \{ g_{j,x} \}_{j=1}^m \rangle$
and $G_Z = \langle \{ g_{j,z} \}_{j=1}^m \rangle$, each of which has
weight less than the maximum.  Contradiction.
Thus we have also shown
$\eta(G) \geq \mbox{max}\{\eta(G_x),\eta(G_z)\}$.
\hfill $\Box$

\subsection{Topological orders from surface cellulations}

This section considers $\delta(G)$ and also $\eta(G)$
in the case in which the stabilizer code $G$ results from the
cellulation of a surface without boundary 
\cite{FreedmanMeyer,bombindelgado}.
We will not review the theory of cellulations or their duals,
except to note that the dual cellulation associates a vertex
to each face of the original and a face to each vertex
(E.g. \cite{Hatcher}).  The relevant definitions will imply that
$\delta(G)$ is the number of edges in the smallest
bounding chain in either the cellulation or its dual.  We also
provide a counterexample to the conjecture that $\delta(G)$ is
the minimum of the valences of the one-skeleton and dual one-skeleton,
although this is frequently the case in examples.

Let $S$ be an oriented surface with no boundary, and let
$\Gamma$ be a two-complex which is a cellulation of $S$.
Let $\mathcal{V}(\Gamma)$, $\mathcal{E}(\Gamma)$, and
$\mathcal{F}(\Gamma)$ denote the vertices, edges, and faces
of $\Gamma$ respectively.  We also suppose a dual cellulation
$\Gamma^\ast$ with bijections
$\mathcal{V}(\Gamma^\ast) \leftrightarrow \mathcal{F}(\Gamma)$,
$\mathcal{E}(\Gamma^\ast) \leftrightarrow \mathcal{E}(\Gamma)$,
and $\mathcal{F}(\Gamma^\ast) \leftrightarrow \mathcal{V}(\Gamma)$
(E.g. \cite{Hatcher}).

Consider the quantum system which associates a qubit to
each edge of $\Gamma$ (or $\Gamma^\ast$).  A well known topologically
ordered stabilizer code has a code space whose dimension is
$\mbox{dim}_{\mathbb{F}_2} H_1(\Gamma,\mathbb{F}_2)$,
where the latter is a cellular homology with bit coefficients
\cite{FreedmanMeyer,bombindelgado}.  To review this briefly, the generators
are indexed by the unions of the faces and vertices of
$\Gamma$.  Let $q(e)$ denote the qubit of an edge 
$e \in \mathcal{E}(\Gamma)$ and $\sigma_{x,q}$ denote the
Pauli tensor which is an identity except for a single $\sigma_x$
factor on qubit $q$.  Then the generator associated to a face
$f \in \mathcal{F}$ is 
$\sigma_x^{\otimes f} = \prod_{e \in \partial f} \sigma_{x,q(e)}$.
The generator associated to a vertex $v \in \mathcal{V}(\Gamma)$
may be defined in terms of the dual face
$v^\ast \in \mathcal{F}(\Gamma^\ast)$.  Namely, let
$\sigma_z^{\otimes v} = \prod_{e^\ast \in \partial v^\ast}
\sigma_{z,q(e)}$, or equivalently tensor over all qubits on
edges incident on the vertex. The stabilizer code is then
$G=\langle \{ \sigma_x^{\otimes f} \}_{f \in \mathcal{F}(\Gamma)} \cup
\{\sigma_z^{\otimes v} \}_{v \in \mathcal{V}(\Gamma)} \rangle$.
Such a code is $XZ$-split.  As an aside, the associated Hamiltonian 
$H=-\sum_{f \in \mathcal{F}(\Gamma)} \sigma_x^{\otimes f}
- \sum_{v \in \mathcal{V}(\Gamma)} \sigma_z^{\otimes v}$
is of interest independent of its homologically structured
degenerate groundstate, in that the excitations out of this
groundstate are abelian anyons with
$\mathbb{Z}/2\mathbb{Z}$ gauge \cite{Kitaev}.

Before considering the topological order as an $XZ$-split stabilizer
code, we set the following notation for homological boundary operators.
\begin{equation}
\begin{array}{llcl}
\partial_2: & \mbox{span}_{\mathbb{F}_2} \mathcal{F}(\Gamma) & \rightarrow
& \mbox{span}_{\mathbb{F}_2} \mathcal{E}(\Gamma), \\
\partial_2^\ast: & \mbox{span}_{\mathbb{F}_2} \mathcal{F}(\Gamma^\ast) 
& \rightarrow
& \mbox{span}_{\mathbb{F}_2} \mathcal{E}(\Gamma^\ast). \\
\end{array}
\end{equation}
Consider matrices $D_X$ and $D_Z$ for $\partial_2$ and
$\partial_2^\ast$ respectively.  Consider a column of $D_X$.
It contains entries of $1 \in \mathbb{F}_2$ at precisely those positions
corresponding to edges $e \in \mathcal{E}$ such that $e \in
\partial_2 f$ for $f \in \mathcal{F}$ the column label.
A similar comment applies to $D_Z$, so that the stabilizer
check matrix of $G$ has this form:
\begin{equation}
A \ = \ \left( \begin{array}{cc}
D_X^T & {\bf 0} \\ 
 {\bf 0} & D_Z^T \\
\end{array} \right).
\end{equation}
Here, the superscript $T$ denotes transpose.  
Also, we list face operators before vertex operators
when forming the matrix, else an antidiagonal matrix results.
Thus, in the special case of a topological order, it is possible to compute
$\delta(G)$ and $\eta(G)$ using only homological inputs, namely
the matrices of the appropriate boundary maps in the cellulation
and cocellulation.

However, $\delta(G)$ and $\eta(G)$ clearly depend on the cellulation
rather than the topology of the underlying surface.  To emphasize
that point, note that for $g \in G_x$ we may associate
$|\mbox{supp}(g)|$ to the size of a boundary in 
$\mbox{span}_{\mathbb{F}_2} \mathcal{E}(\Gamma)$, while a similar
comment applies to $g_z$ and 
$\mbox{span}_{\mathbb{F}_2} \mathcal{E}(\Gamma^\ast)$.
Hence, $\delta(G)$ is the minimum of the smallest number of
edges required to support a boundary in either $\Gamma$ or
$\Gamma^\ast$.  Since one may always subdivide an edge, this
is not a topological invariant.

It is tempting given the last paragraph to conjecture that
$\delta(G)$ is the minimum of the valences of the one-skeleta
of $\Gamma$ and $\Gamma^\ast$, i.e. of the graphs which result
by ignoring faces (two-cells) in either.  In fact, this is incorrect.
Figure \ref{fig:1} provide a counterexample, 
in that the boundary
with the least number of edges in $\Gamma$ does not bound
a single face.

\section{Conclusions}

Since the locality of a Hermitian
matrix might serve as a crude figure-of-merit for 
its experimental difficulty, theorists
hope to find interactions which are both highly local and have
robust stabilizer quantum codes as their groundstate eigenspaces
(E.g. \cite{dur}).
This \dpo{manuscript} constrains such efforts by arguing
that excessively local effective Hamiltonians either do not exist
or else have undesirable properties.  The three main results
argue that (i) no effective Hamiltonian may be more local than
the minimum locality of any element of the stabilizer group,
(ii) effective Hamiltonians which do not allow for more
nonlocal Pauli tensors in the stabilizer group must also
be effective Hamiltonians for many other stabilizer groups,
and (iii) approximating a stabilizer code using an excessively
local Hamiltonian leads to gap pinching.  Nonetheless, the
technical statements given here might lead to new examples.

Speculating in a slightly broader context, there are 
two ways one might attempt to use 
$k$-local Hamiltonians to simulate $\ell>k$-local systems of interest.
One might 
(a) exploit crosstalk mediated by an ancilla or (b) pulse
noncommuting Hamiltonians (absent ancilla).
Our results do not account for clever use
of ancilla.  Indeed, ancillae have been used successfully
to construct effective Hamiltonians in the adiabatic 
computing literature \cite{gadget}.
On the other hand,
our results argue that Hamiltonians whose addends are 
\emph{noncommuting} Pauli tensors must nonetheless obey
certain locality constraints on their groundspaces.  Thus,
the ancilla-based approach might be preferable.

\end{document}